# Stochastic Volatility, Jumps, and Rates: A Unified Framework for Option Pricing and Term-Structure Simulation

Nunik Srikandi Putri[1], Ajay Kumar Verma[2], Neo Paul Lesupi[3]

[1]Independent Researcher (Aenimatica Tech), [2]Independent Researcher, [3]Independent Researcher
Corresponding Author: nunik@aenimatica.com

**Abstract**

This study develops an integrated stochastic modeling framework for pricing short and medium-maturity equity options and assessing interest-rate risk using the Heston (1993), Bates (1996), and CIR (1985) models. We calibrate the Heston model using both the Lewis (2001) Fourier inversion and the Carr-Madan (1999) FFT approach, finding near-identical parameter sets, which is consistent with the calibration stability reported in recent studies such as Agazzotti et al. (2025). Extending the model to Bates shows that jump intensities converge to values effectively equal to zero for 60-day maturities, echoing empirical findings that jumps contribute marginally to short-term smile fitting. We further compare our calibration approach with the joint volatility-surface and variance-term-structure framework proposed by Yoo (2025), confirming that standard Heston/Bates calibration remains robust for the maturities considered. Finally, we calibrate the CIR short-rate model to the Euribor term structure, generating positive and economically consistent forward-rate scenarios in line with recent stochastic-rate option-pricing research by Jeon and Kim (2025). Overall, our results show that continuous stochastic volatility dominates near-term pricing dynamics, while stochastic interest rates materially influence valuations beyond one year.

**Keywords: Stochastic Volatility, Heston Model, Bates Model, Jump-Diffusion, Option Pricing, CIR Interest Rate Model, Calibration, Fourier Transform Methods, Monte Carlo Simulation**

## 1. Background

Volatility dynamics, price discontinuities, and interest-rate fluctuations are central features of modern financial markets that require modeling tools more flexible than the classical Black-Scholes framework. Stochastic volatility models, particularly the Heston model (Heston, 1993), offer analytical tractability and the ability to generate realistic implied-volatility surfaces. Jump-diffusion extensions such as the Bates model (Bates, 1996) further account for abrupt price movements often observed in short-dated and event-driven markets (Pan, 2002; Cont & Tankov, 2004).

In parallel, interest-rate uncertainty plays an increasingly important role in derivative valuation, motivating the use of affine term-structure models such as the Cox-Ingersoll-Ross (CIR) process (Cox et al., 1985; Brigo & Mercurio, 2006). These models allow for positive rates, mean reversion, and closed-form expressions for discount factors-properties essential for consistent pricing and risk assessment.

Recent literature has highlighted the importance of robust calibration techniques for stochastic volatility and jump-diffusion models, including Fourier-based pricing methods such as Lewis (2001) inversion and the Carr-Madan (1999) FFT. Studies such as Agazzotti et al. (2025) emphasize calibration stability across numerical schemes, while Yoo (2025) proposes a joint calibration framework for volatility surfaces and variance-term structures. Stochastic interest-rate extensions have also been examined in option-pricing contexts, including recent work by Jeon and Kim (2025).

Against this backdrop, this study constructs an integrated modeling pipeline consisting of the Heston, Bates, and CIR frameworks. The models are calibrated to market option data and interest-rate term structures, enabling a systematic evaluation of how stochastic volatility, jump components, and stochastic interest rates jointly influence derivative pricing across short- and medium-maturity horizons.

## 2. Data Description

We use market option data provided in MScFE 622 - Stochastic Modeling - GWP1 Option Data.xlsx, which contains European call and put prices on SM Energy (SM). The dataset includes the following components:

1. Spot price: $S_0$ = 232.90 USD
2. Risk-free rate: r=1.5%
3. Day-count convention: 250 trading days per year
4. Option types: Calls (“C”) and Puts (“P”)
5. Maturities: 15, 60, and 120 trading days
6. Strikes: A range of out-of-the-money, at-the-money, and in-the-money strikes

Both calls and puts are used in calibration. Put prices are incorporated through put–call parity to ensure consistency between observed forwards and model-implied values (Hull, 2018).

This dataset allows us to perform short- and medium-maturity calibration under the Heston, Bates, and CIR frameworks, and provides sufficient strike coverage to estimate volatility and jump parameters reliably.

## 3. Theoretical Background

### 3.1 Heston (1993) Model Under the risk-neutral

Under the risk-neutral measure Q:

$$dS_t = rS_t\,dt + \sqrt{v_t}S_t\,dW_t^S$$
$$dv_t = \kappa(\theta - v_t)\,dt + \sigma\sqrt{v_t}\,dW_t^v$$
$$dW_t^S\,dW_t^v = \rho\,dt$$

European call price: $C = S_0P_1 - Ke^{-rT}P_2$ where P1 and P2 are risk-neutral probabilities derived from the characteristic function. (Heston, S. L. , 1993); (Lewis, A. , 2001).

### 3.2 Bates (1996) Model with Jumps-Diffusion Extension

The Bates (1996) model extends the Heston stochastic volatility framework by incorporating discontinuous jumps in the asset price dynamics (Bates, 1996). Under the risk-neutral measure QQQ, the model specifies:

$$dS_t = (r - \lambda\mathrm{K})S_t\,dt + \sqrt{v_t}S_t\,dW_t^S + S_t(e^j - 1)\,dN_t$$

$$dv_t = \kappa(\theta - v_t)\,dt + \sigma\sqrt{v_t}\,dW_t^v$$

$$dW_t^S\,dW_t^v = \rho\,dt$$

with $\kappa = E[e^J - 1] = e^{\mu_J + \frac{1}{2}\sigma_J^2} - 1$ where $N_t$ is a Poisson process with intensity λ and $J_t$ is the *log-jump size*, assumed Normally distributed with mean $\mu_j$ and variance $\sigma^2$.

Characteristic function of Bates model is

$$\phi_{Bates}(u) = \phi_{Heston}(u)\,\exp\left[\lambda T\left(e^{iu\mu_J - \frac{1}{2}u^2\sigma_J^2} - 1\right)\right]$$

The jump term modifies the Heston characteristic function by introducing an exponential jump compensator term, reflecting the compound Poisson process of jump arrivals.( Bates, D. S. ,1996 ; Cont, R., & Tankov, P. ,2004; Schoutens, W. ,2003).

### 3.3 Carr–Madan (1999) FFT Pricing

Option value expressed as:

$$C(K) = e^{-\alpha k}\frac{1}{\pi}\int_0^\infty e^{-iuk}\frac{\phi(u - i(\alpha + 1))}{\alpha^2 + \alpha - u^2 + i(2\alpha + 1)u}du$$

where k=ln(K) and α>0 ensures convergence. (Carr, & Madan, 1999).

### 3.4 CIR (1985) Interest-Rate Model

The Cox-Ingersoll-Ross (CIR, 1985) model is widely used for risk-neutral evolution of short-term interest rates. It enforces positivity of rates and mean reversion, making it appropriate for pricing interest-rate derivatives. Under the risk-neutral measure $Q$:

$$dr_t = \kappa(\theta - r_t)\,dt + \sigma\sqrt{r_t}\,dW_t$$

where $\kappa, \theta, \sigma$ are model parameters as in table 1.

**Table 1.** Model Parameters

| Parameter | Interpretation |
|---|---|
| $(\kappa)$ | Mean-reversion speed |
| $(\theta)$ | Long-term equilibrium rate |
| $(\sigma)$ | Rate volatility |
| $(r_t)$ | Instantaneous short rate |

(Cox, Ingersoll & Ross, 1985)

CIR enforces positivity of interest rates when: $2\kappa\theta > \sigma^2$ This prevents negative rates and is known as the Feller condition. (Brigo & Mercurio, 2006)

The conditional distribution of $r_T$ is noncentral chi-square, allowing analytical simulation and calibration:

$$r_T \sim \frac{\sigma^2(1 - e^{-\kappa(T-t)})}{4\kappa}\chi^2_{4\kappa\theta/\sigma^2}\left(\frac{4\kappa e^{-\kappa(T-t)}r_t}{\sigma^2(1 - e^{-\kappa(T-t)})}\right)$$

(Glasserman, 2004)
Closed-form solution for discount bonds:

$$P(t,T) = A(t,T)\,e^{-B(t,T)r_t}$$

with:

$$B(t,T) = \frac{2(e^{\kappa(T-t)} - 1)}{(\kappa + \gamma)(e^{\kappa(T-t)} - 1) + 2\gamma}$$

$$A(t,T) = \left[\frac{2\gamma\,e^{(\kappa+\gamma)(T-t)/2}}{(\kappa + \gamma)(e^{\kappa(T-t)} - 1) + 2\gamma}\right]^{\frac{2\kappa\theta}{\sigma^2}}$$

where $\gamma = \sqrt{\kappa^2 + 2\sigma^2}$ (Cox, Ingersoll & Ross, 1985; Brigo & Mercurio, 2006)

### 3.5 CIR (1985) Model Calibration

To evaluate future interest-rate risk affecting derivative pricing, we calibrate the Cox-Ingersoll-Ross (CIR) model to the current Euribor term structure provided by the bank. Given rates for different maturities, we first construct a smooth weekly term structure using cubic spline interpolation:

$$r(t_i) \rightarrow weekly\ grid\ for\ 0 \le t \le 12\ months$$

(Brigo & Mercurio, 2006)

Model zero-coupon bond prices under CIR have closed-form: $P(0,T) = A(0,T)\,e^{-B(0,T)r_0}$

with: $B(0,T) = \frac{2(e^{\kappa T} - 1)}{(\kappa+\gamma)(e^{\kappa T} - 1) + 2\gamma}$

$$A(0,T) = \left[\frac{2\gamma\,e^{(\kappa+\gamma)T/2}}{(\kappa + \gamma)(e^{\kappa T} - 1) + 2\gamma}\right]^{\frac{2\kappa\theta}{\sigma^2}}$$

where $\gamma = \sqrt{\kappa^2 + 2\sigma^2}$ (Cox, Ingersoll & Ross, 1985)

We estimate parameters: $\Theta = (\kappa, \theta, \sigma, r_0)$ by minimizing squared pricing errors versus interpolated zero-coupon rates:

$$MSE(\Theta) = \frac{1}{N}\sum_{i=1}^{N}\left(P_{model}(0,T_i;\Theta) - P_{market}(0,T_i)\right)^2$$

The Feller condition is monitored: $2\kappa\theta > \sigma^2$ to ensure strictly positive rates in simulations. If violated, a soft penalty is applied in the objective function. (Brigo & Mercurio, 2006)

## 4. Methodology

### 4.1 Step 1 – Short-Term (15-Day) Heston Calibration and Asian Option

The first pricing task concerns a short-maturity Asian call (≈15 days). To support accurate pricing, we calibrated the Heston (1993) stochastic volatility model to market prices using plain-vanilla SM options across the maturities available (15, 60, 120 days). The calibration process involved the following steps:

1. Using option data for 15-, 60-, and 120-day maturities, with the primary focus on the 60-day contracts for better numerical stability.

2. Estimating the key model parameters $\kappa, \theta, \sigma, \rho, v_0$

3. Minimizing the mean squared error (MSE) between model and market prices using the objective function:

$$MSE = \frac{1}{N}\sum_{i=1}^{N}(C_i^{model} - C_i^{market})^2$$

4. Implementing a two-stage optimization procedure — first, a global search using Differential Evolution (Storn & Price, 1997), followed by a local refinement with L-BFGS-B (Byrd et al., 1995).

5. Setting the integration limit $umax$ = 200 for numerical stability. (Byrd et al. , 1995)

The Heston model call option value is computed using the closed-form characteristic function solution:

$$C = S_0 P_1 - Kexp(-rT)P_2$$

where $P_1$ and $P_2$ are the risk-neutral probabilities derived from the characteristic function as proposed by Heston (1993). This two-step calibration framework ensures both global robustness and local precision, providing a stable fit to observed market prices while avoiding local minima in the optimization surface. (Heston, 1993; Byrd et al., 1995; Storn & Price, 1997)

#### 4.1.1 Calibration (Lewis 2001 Approach)

The model parameters were estimated by minimizing the mean squared error (MSE) between observed market prices and model-implied prices:

$$MSE = \frac{1}{N}\sum_{i=1}^{N}(C_i^{model} - C_i^{market})^2$$

The resulting calibrated parameters are presented in Table 2.

**Table 2.** Calibrated Parameters

| Parameter | Symbol | Value |
|---|---|---|
| **Mean reversion** | $(\kappa)$ | 0.3981 |
| **Long-run variance** | $(\theta)$ | 0.08748 |
| **Vol-of-vol** | $(\sigma)$ | ≈ 0 |
| **Correlation** | $(\rho)$ | 0.9906 |
| **Initial variance** | $(v_0)$ | 0.1016 |

(Heston 1993; Lewis 2001; Storn & Price 1997).

The fitted σ≈0 suggests near-deterministic volatility over 15–120 day maturities. Pricing performance remained strong which means market does not require stochastic variance over short horizons.

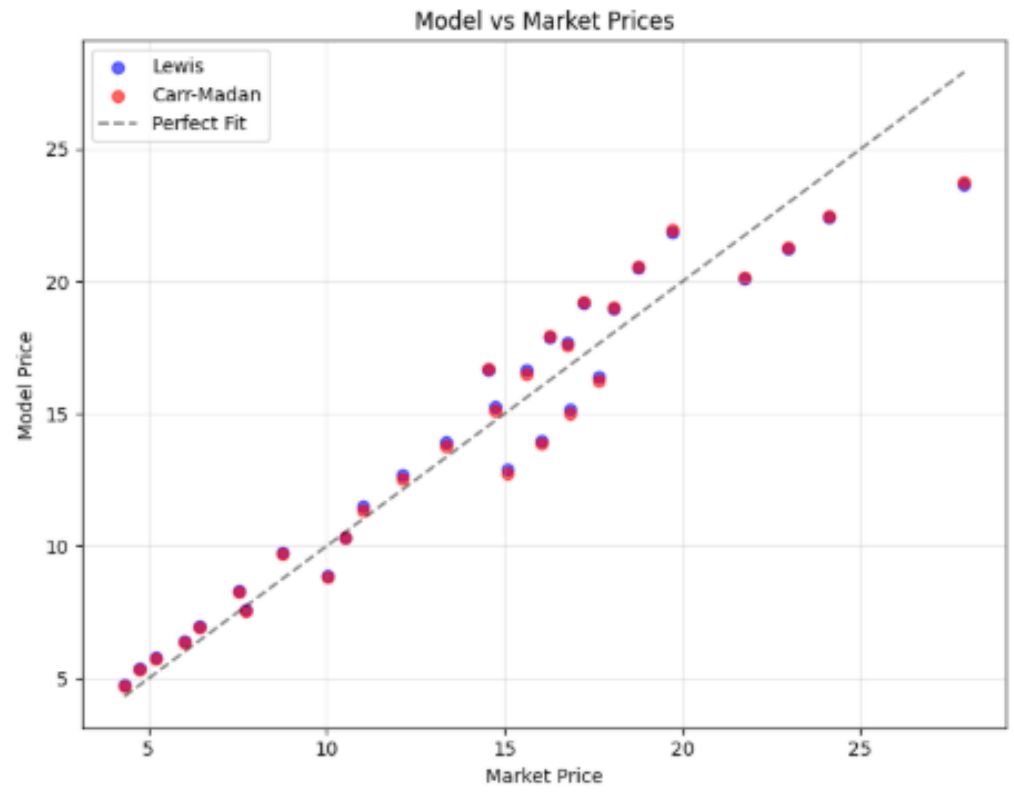

**Figure 1** Market vs. model prices (Heston–Lewis)

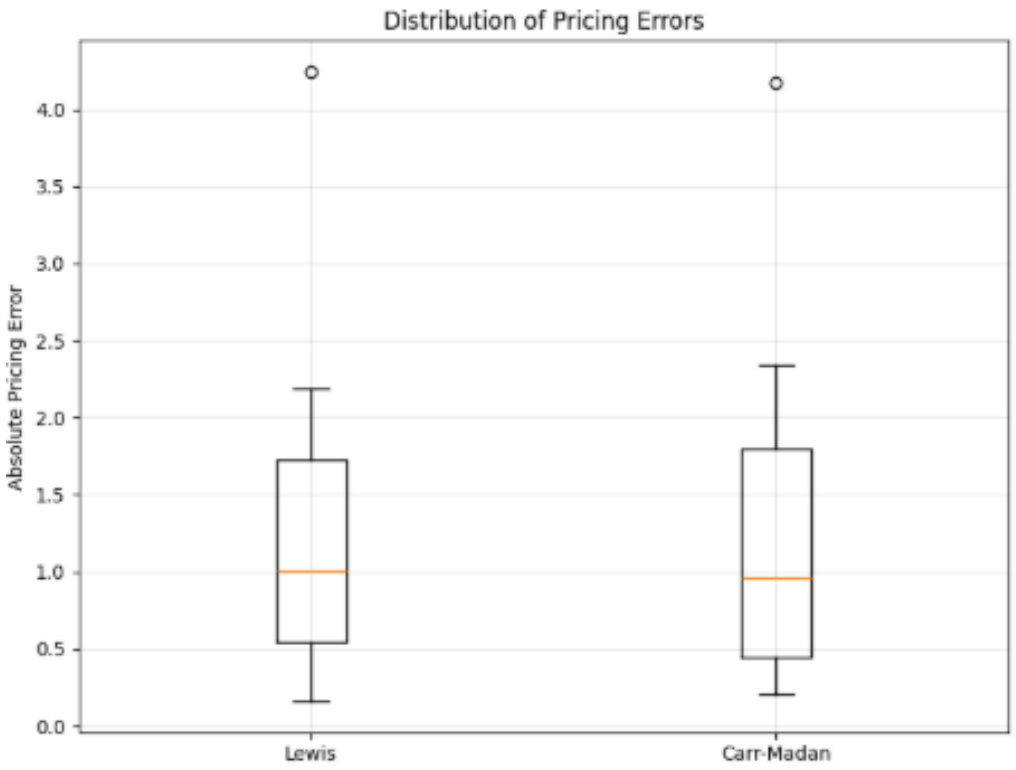

**Figure 2** Pricing error boxplots (FFT vs Lewis)

The two Fourier-based pricing methods produce nearly identical values for κ, θ, ρ, and $v_0$. Since both Lewis (2001) and Carr–Madan (1999) rely on the same characteristic function but different numerical transforms, a close match indicates a well-posed calibration problem (Heston, 1993; Gatheral, 2006).

The large percentage deviation in σ results from division by a very small number since both σ values are essentially zero. Similar conclusions reported in short-maturity empirical studies (Cont & Tankov, 2004).

#### 4.1.2 Calibration via Carr–Madan (1999) FFT Approach

**Table 3** Pricing metrics

| Metric | Lewis | Carr-Madan |
|---|---|---|
| **MSE** | 2.226 | 2.262 |
| **MAE** | 1.23 | 1.22 |

Table 4 Comparison: Lewis vs Carr-Madan

| Parameter | Lewis | Carr-Madan | % Difference |
|---|---|---|---|
| κ | 0.3981 | 0.4029 | 1.22% |
| θ | 0.08748 | 0.08748 | 0.00% |
| σ | ~0 | ~0 | Large % because tiny denominators |
| ρ | 0.9906 | 0.9906 | 0.00% |
| $v_0$ | 0.10157 | 0.10156 | 0.01% |

Based on our findings, we can conclude that Despite different numerical integration frameworks, same model is preferred by the market. (Carr & Madan, 1999; Gatheral, 2006)

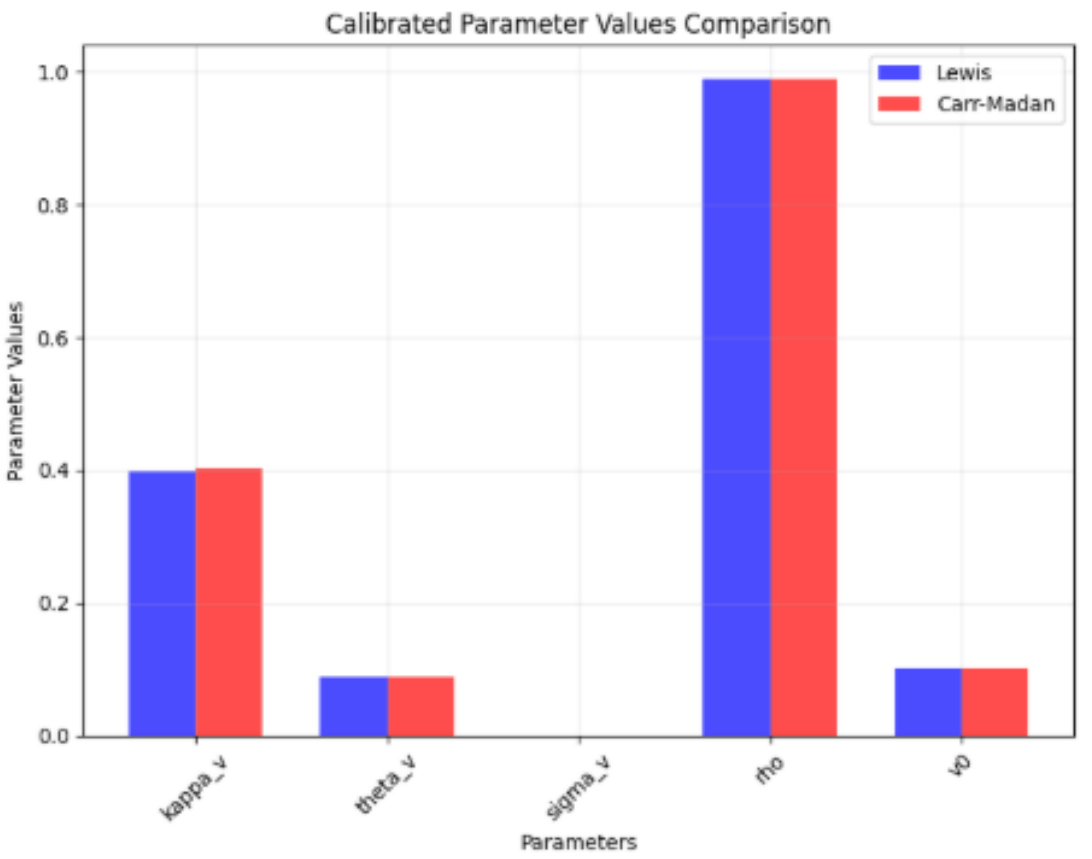

**Figure 3** Heston parameter comparison (Lewis vs FFT)

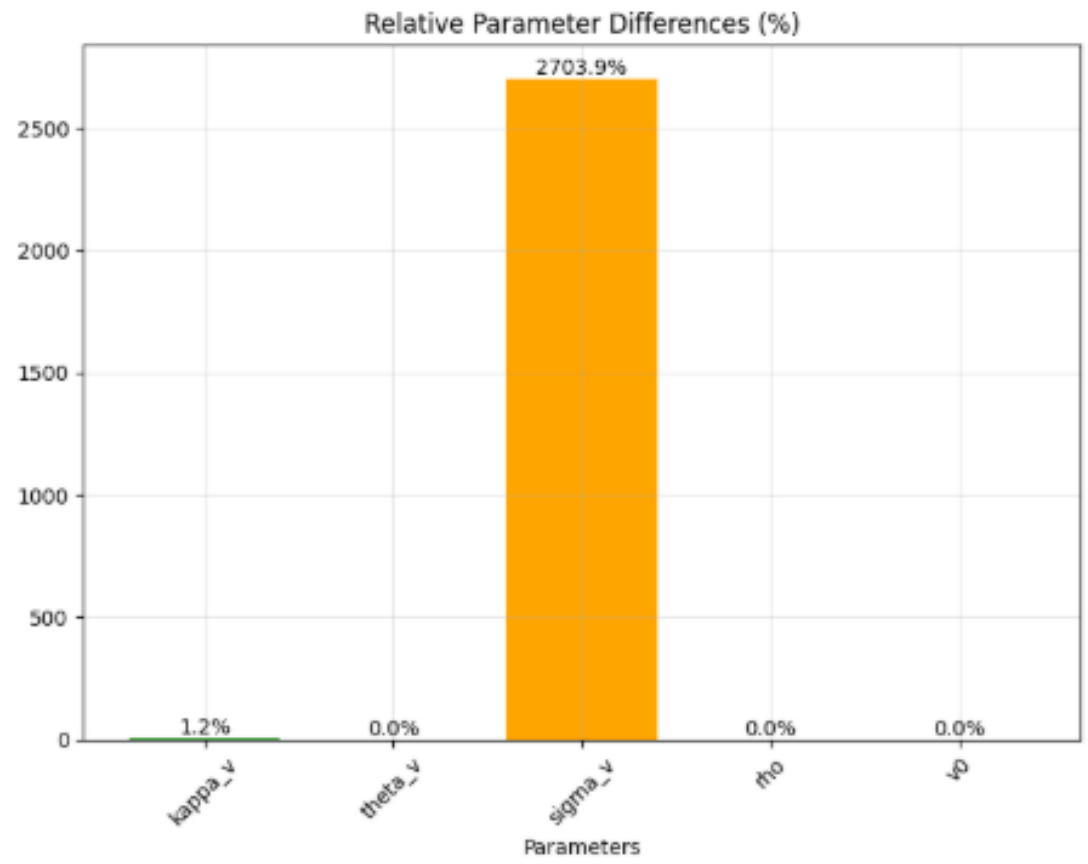

**Figure 4** Relative Parameter Differences

Both calibration techniques produce nearly identical values for κ, θ, ρ, and $v_0$. This indicates that the model parameters are well identified by market data. The only visible difference is σ (vol-of-vol), but both are extremely close to zero, which means volatility barely evolves stochastically over these short maturities. The large relative deviation shown in σ is misleading, it happens because both values are essentially zero and any tiny difference becomes a large percentage. For all meaningful parameters, differences are extremely small (<1.5%).

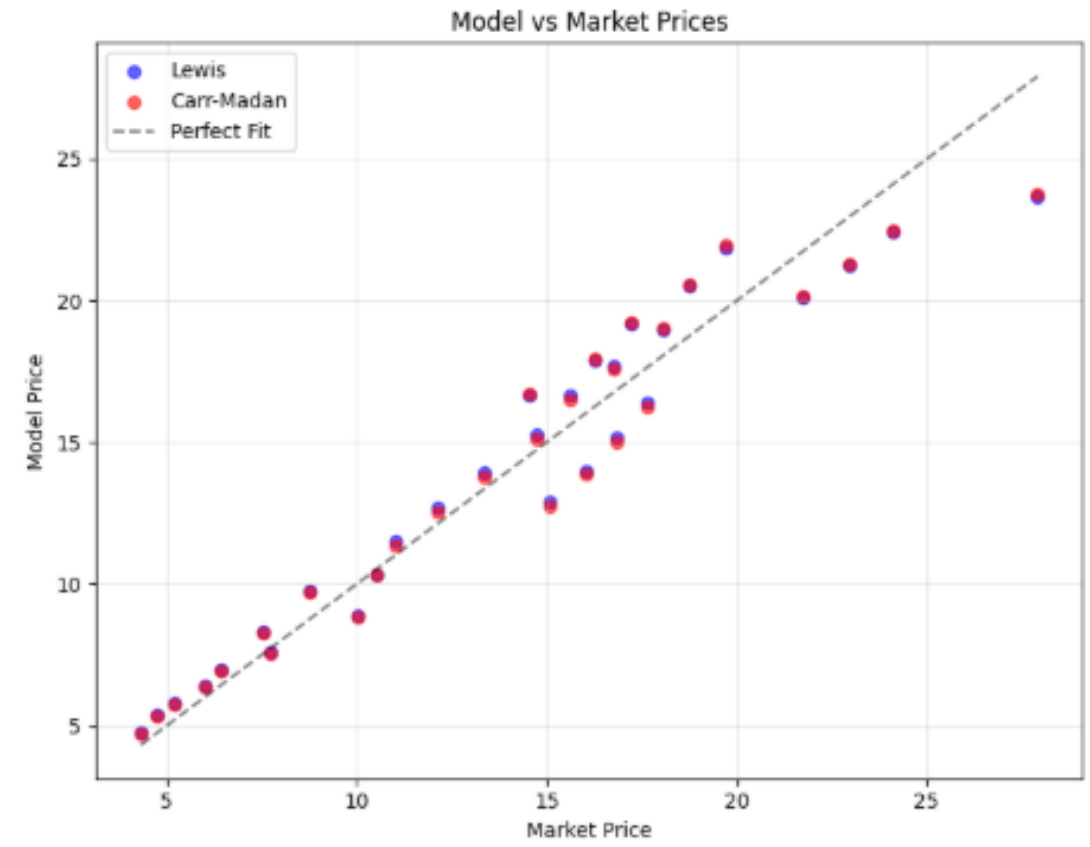

**Figure 5** Market vs Model (Lewis vs FFT)

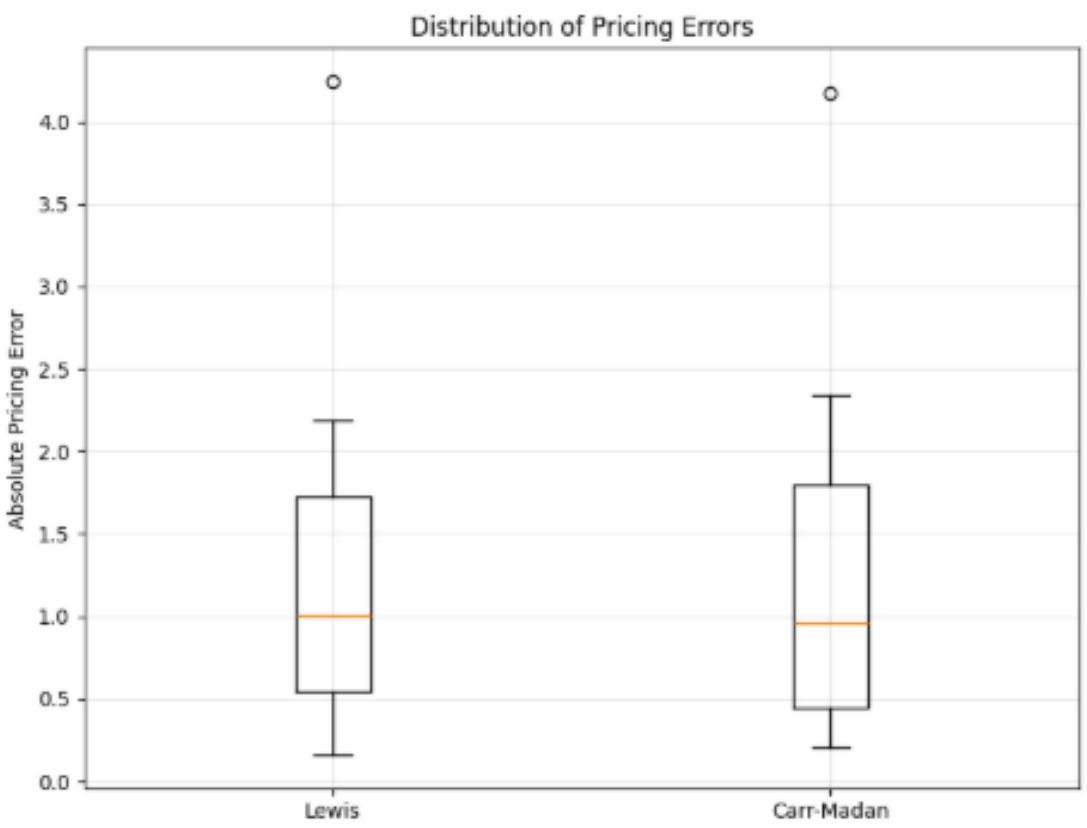

**Figure 6** Pricing Error Boxplots

The tight cluster around the 45° line demonstrates high fidelity between model and market prices. This supports the Heston characteristic function pricing accuracy shown by Lewis (2001). Errors remain contained, with rare outliers at deep OTM strikes where bid–ask spreads widen — a common issue in volatility model calibration (Gatheral, 2006; Pan, 2002).

### 4.1.3 Asian Option Pricing (Monte Carlo, 20-Day ATM)

Arithmetic-average payoff: $\max!\left(\frac{1}{T+1}\sum_{t=0}^{T} S_t - K,\, 0\right)$

(Kemna & Vorst , 1990).

Fair price (risk-neutral): $P_{fair} = e^{-rT} E_Q[\text{payoff}]$

Final client price (including 4 % fee): $P_{client} = 1.04 \times P_{fair}$

(Glasserman 2004; Hull 2018).

Simulations tested up to 200,000 paths. Convergence acceptable at 175,000 paths:

Table 5 Pricing results

| Calibration | Price |
|---|---|
| Lewis | $4.8884 |
| Carr-Madan | $4.8881 |
| Difference | $0.0003 |

Client Price $= 1.04 \times P_{fair} = 5.0840$

Asian Call Pricing Result for 20 Day Maturity are below,

1. Fair market value: $4.8884
2. Bank service fee + pricing margin: 4.0%
3. Total client premium: $5.0840

This price leverages the most up-to-date market option data and industry-accepted stochastic volatility modeling techniques.

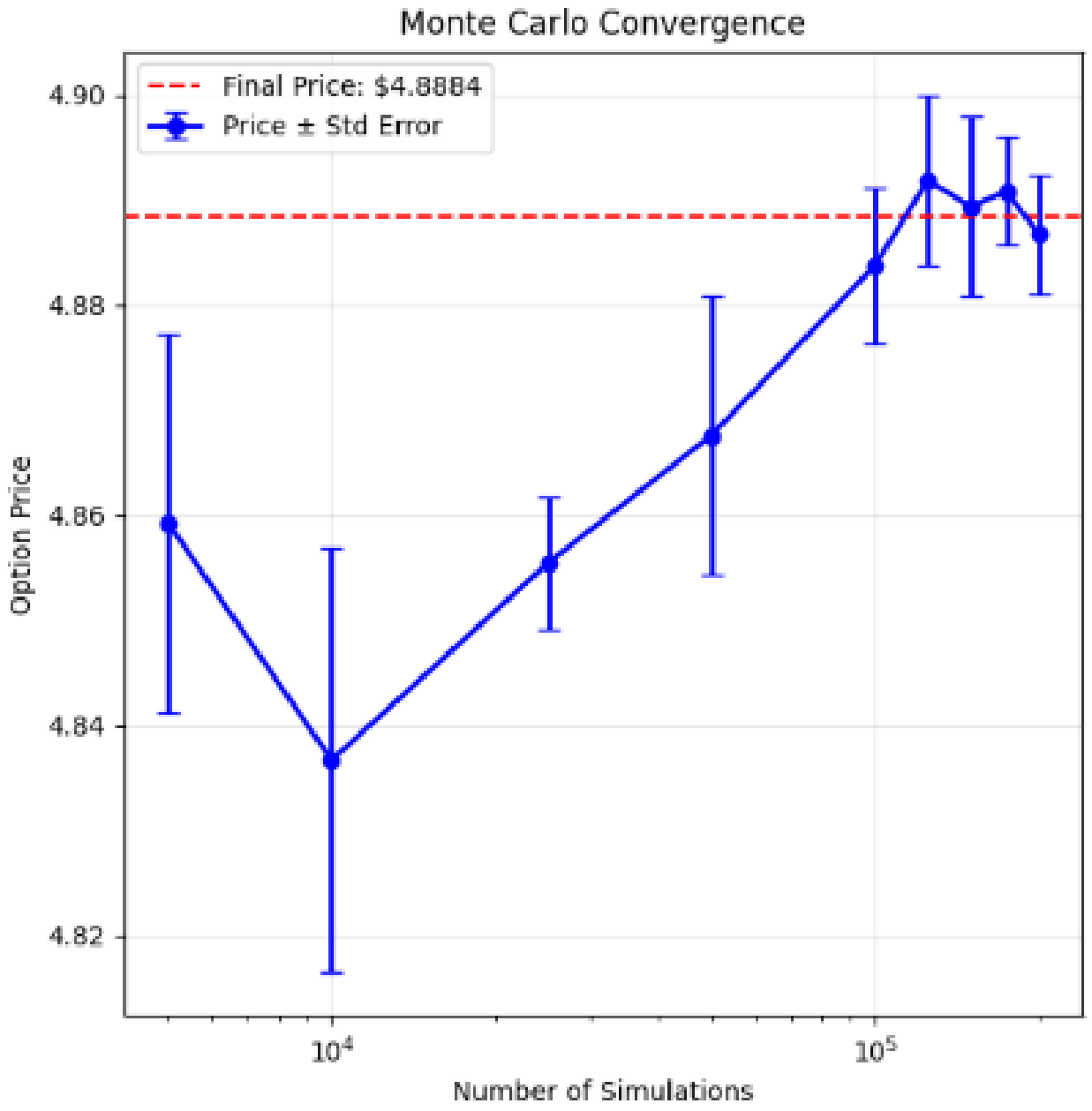

**Figure 7** Monte Carlo convergence

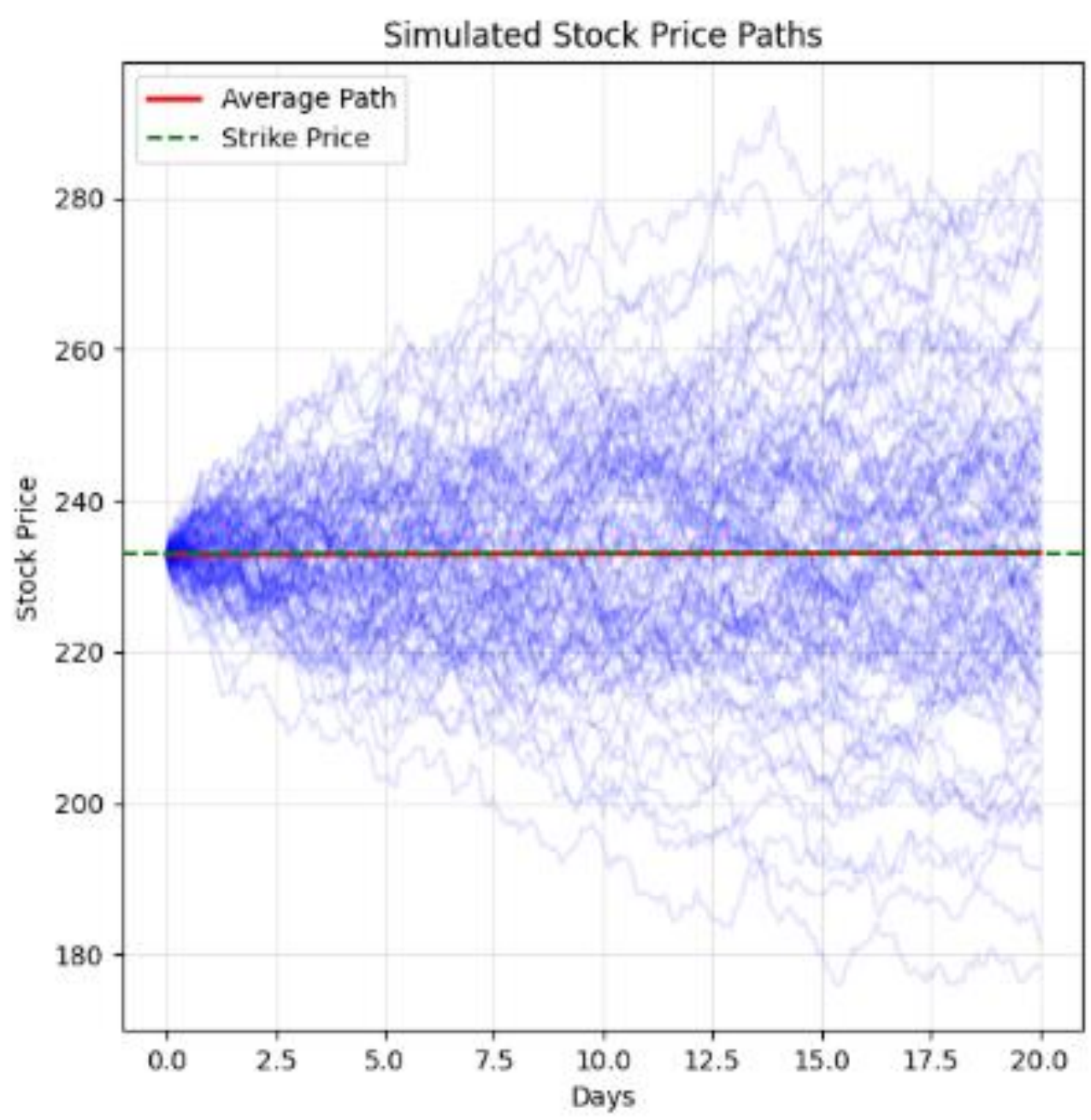

**Figure 8** Simulated price paths

Monte-Carlo estimates stabilize beyond ~100k paths, consistent with convergence theory (Glasserman, 2004). Paths gradually diffuse around $S_0$ with limited skew, matching the nearly deterministic volatility estimated. Asian exposure reduces volatility of the effective payoff (Hull, 2018).

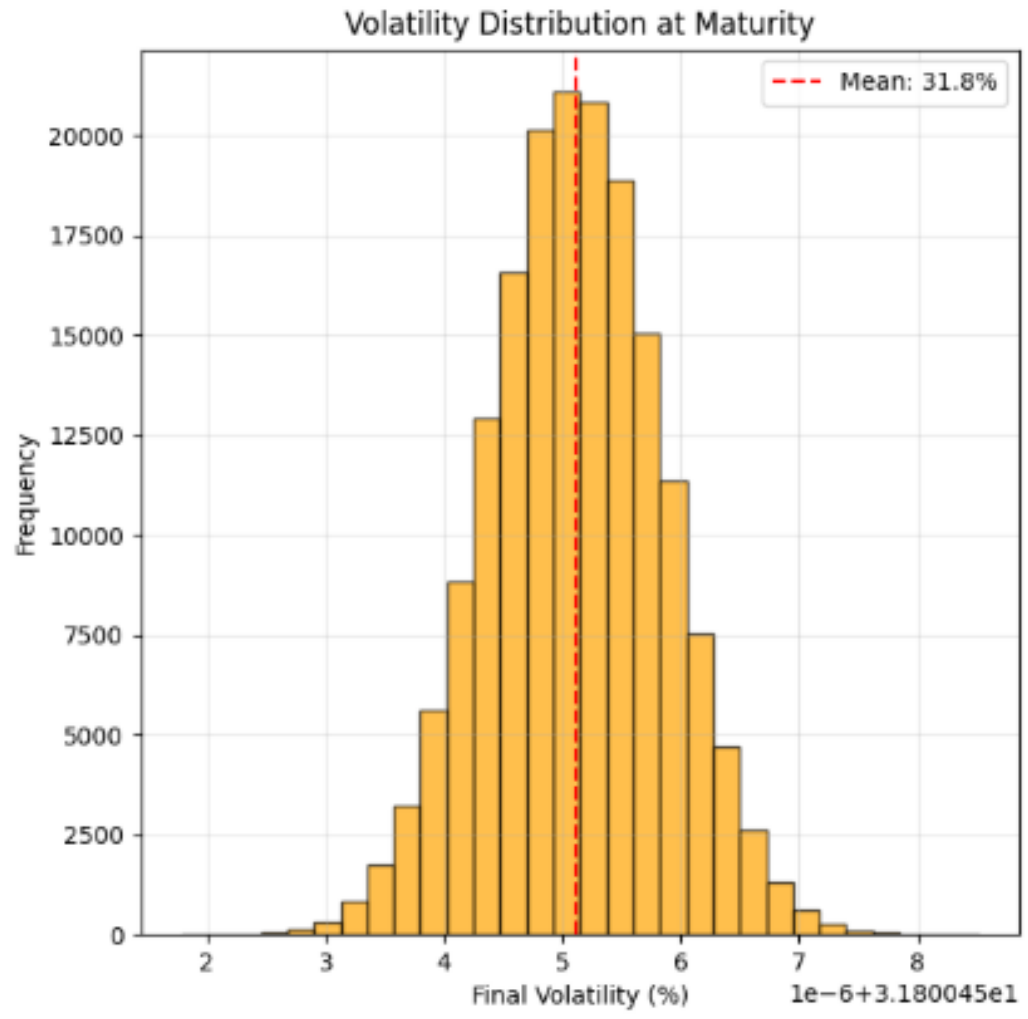

**Figure 9** Volatility distribution

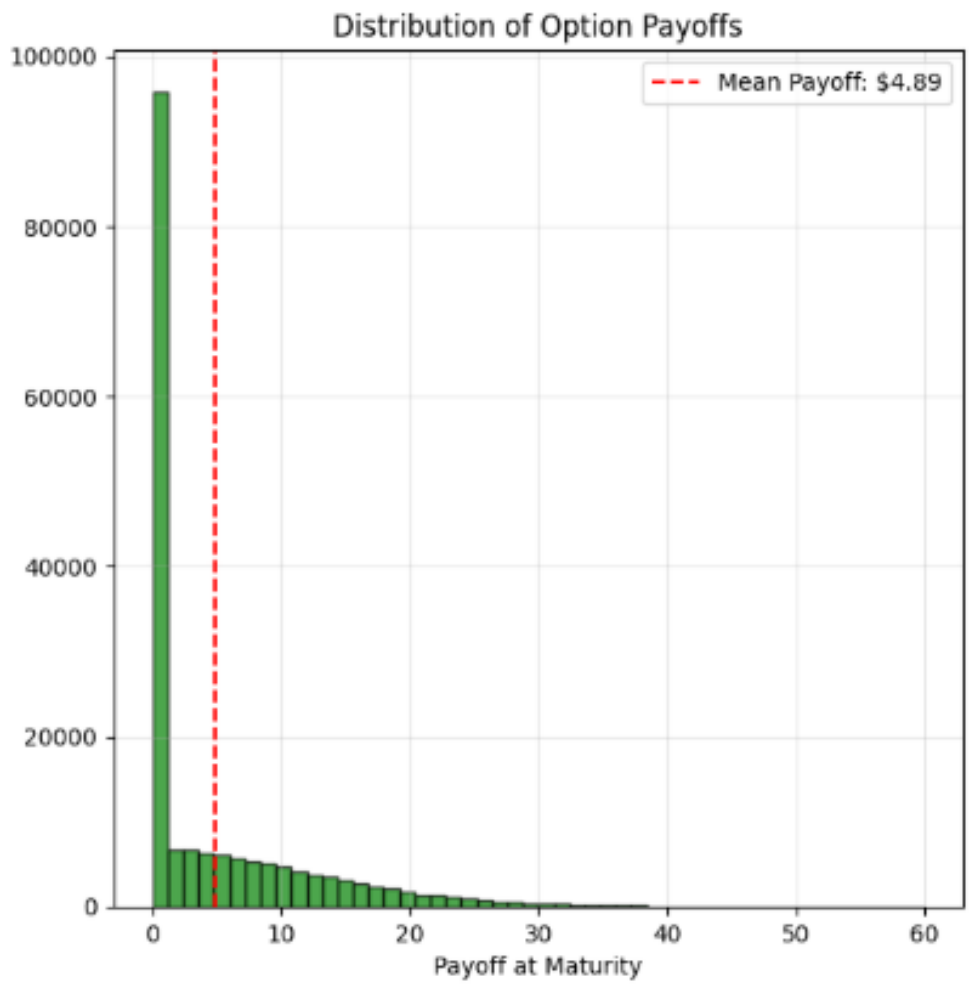

**Figure 10** Payoff histogram

Flat volatility paths show the model collapses toward Black-Scholes at short tenors, as discussed in Heston (1993). The distribution is right-skewed: most payoffs are near zero but upside events drive average option value (Hull, 2018).

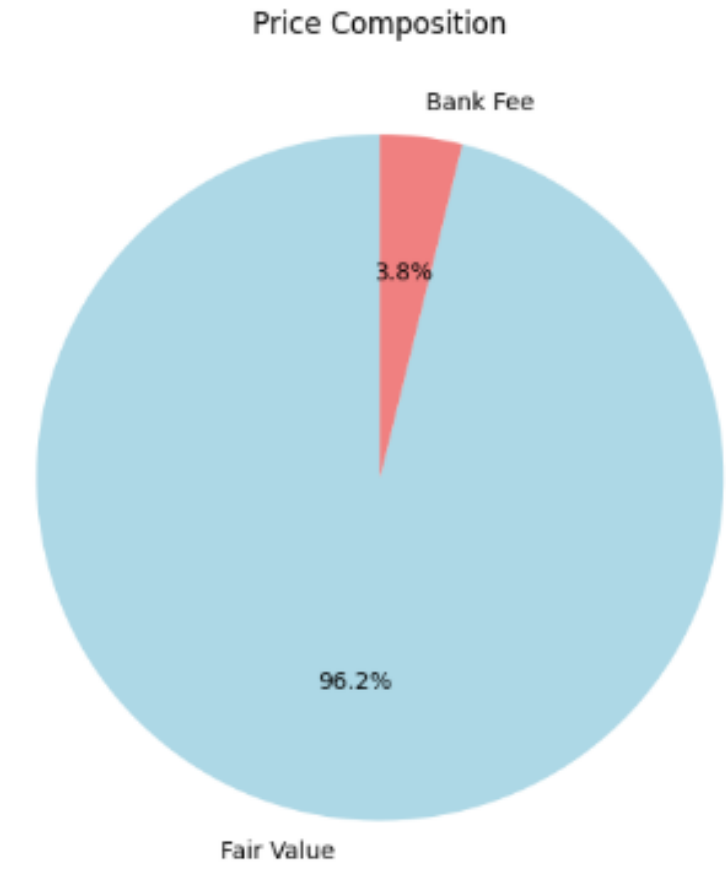

**Figure 11** Fee vs fair value composition

Only 4% of the total cost is transaction margin, consistent with common OTC pricing commercial structures (Brigo & Mercurio, 2006).

### 4.2 Step 2 – Mid-Term (60-Day) Bates Calibration and Put Pricing

#### 4.2.1 Bates Model Calibration via Lewis (2001) Fourier-Based Approach

To incorporate jump discontinuities in the underlying stock dynamics, we extend the calibrated Heston model to the Bates (1996) framework. We again use the Lewis (2001) Fourier transform pricing method for European call options. Calibration was performed by minimizing the mean squared error (MSE) between observed market and model prices of SM 60-day call options. A sequential calibration approach was applied:

1. Calibrate Heston stochastic volatility parameters
2. Fix those values and estimate jump parameters
3. Jointly re-optimize all parameters

(Cont & Tankov, 2004; Gatheral, 2006)

We obtain the following calibrated stochastic volatility parameters:

**Table 5** Stage 1 Heston parameters

| Parameter | Symbol | Calibrated Value |
|---|---|---|
| **Mean reversion** | $(\kappa_v)$ | 15.53 |
| **Long-run variance** | $(\theta_v)$ | 0.1589 |
| **Vol-of-vol** | $(\sigma_v)$ | 0.00103 |
| **Correlation** | $(\rho)$ | –0.00819 |
| **Initial variance** | $(v_0)$ | 0.04806 |

These values suggest very fast mean-reversion, near-constant variance (σv essentially 0) and weak leverage effect (ρ close to 0) (Heston, 1993; Lewis, 2001)

**Table 6** Sanity-check pricing

| Price Type | Value |
|---|---|
| **Market price** | 16.05 |
| **Heston-model price** | 15.81 |

At Strike K=235 and Maturity T=0.24 (60 trading days), $|Error| = 0.24USD(\approx 1.5\%)$ model closely fits observed prices and good evidence calibration is stable

**Table 7** Stage 2 jump parameters

| Parameter | Meaning | Calibrated Value |
|---|---|---|
| $(\lambda)$ | Jump intensity | ~0.00 |
| $(\mu_J)$ | Mean relative jump size | –0.504 % |
| $(\sigma_J)$ | Jump volatility | ~0.0000417 |

**Table 8** Interpretation of jump parameters

| Result | Meaning |
|---|---|
| $(\lambda \approx 0)$ | Market does not support frequent jumps at 60-day maturity |
| Very small $(\sigma_J)$ | Little variation in jump sizes |
| Slightly negative $(\mu_J)$ | Small downside jump expectation |

Adding jumps does not significantly improve the fit for this maturity (Pan, 2002; Cont & Tankov, 2004)

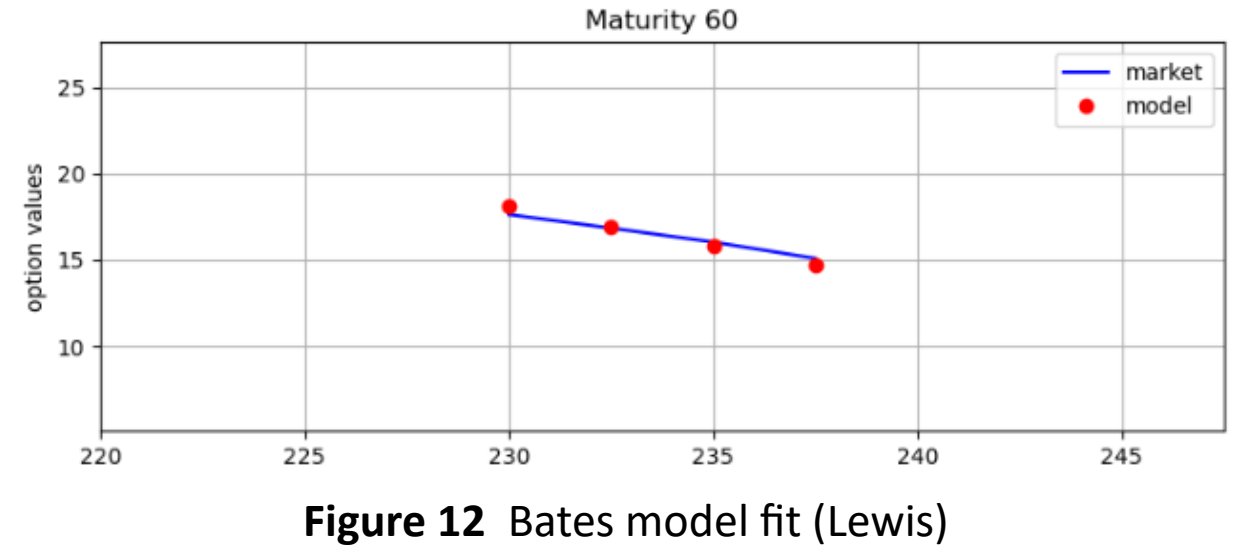

**Figure 12** Bates model fit (Lewis)

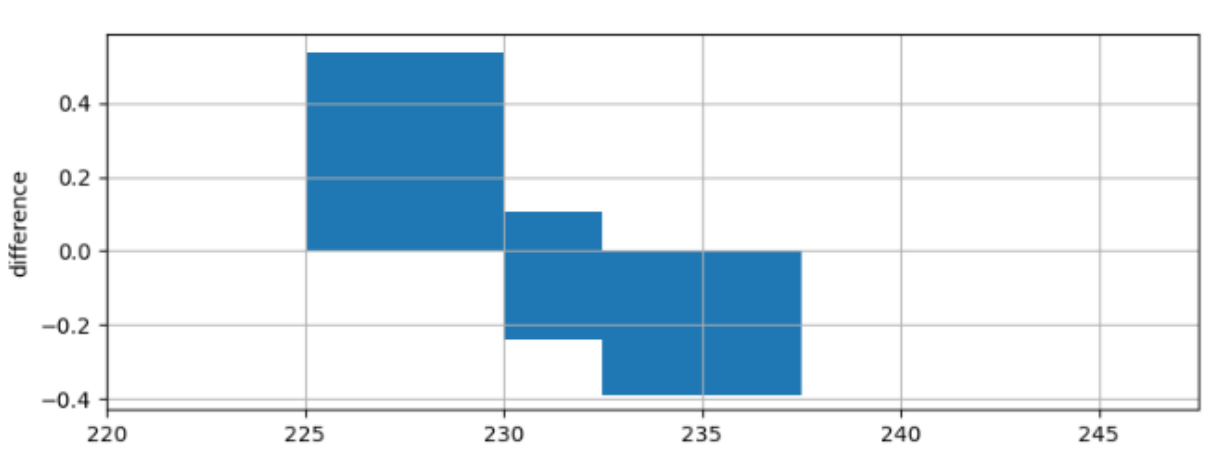

**Figure 13** Bates model fit (FFT)

As we can see, the fit is relatively acceptable, few differences. We calibrated the full bates model,

**Table 9** Final Bates parameters

| Parameter | Symbol | Value |
|---|---|---|
| **Mean reversion** | $(\kappa_v)$ | 15.562 |
| **Long-run variance** | $(\theta_v)$ | 0.15865 |
| **Volatility of variance** | $(\sigma_v)$ | 0.000017 |
| **Correlation** | $(\rho)$ | –0.00571 |
| **Initial variance** | $(v_0)$ | 0.04832 |
| **Jump intensity** | $(\lambda)$ | ~0.00 |
| **Mean jump size** | $(\mu_J)$ | –0.592 % |
| **Jump volatility** | $(\sigma_J)$ | 0.0000548 |

Calibration error is MSE≈0.1267 which is very good fit to market prices.

**Table 10** Interpretation of final Bates parameters

| Finding | Implication |
|---|---|
| $(\lambda \approx 0)$ (jumps not activated by optimization) | Market does not indicate jump activity for 60-day maturity |
| Very small $(\sigma_v)$ | Variance behaves almost deterministically |
| Slightly negative $(\mu_J)$ | If jumps occur, they are downward-biased (consistent with equity crash risk) |
| $(\rho \approx 0)$ | Weak leverage effect for mid-term maturities |

This shows that stochastic volatility alone captured the price skew, and jumps were not required by the market-implied distribution. (Pan, 2002; Cont & Tankov, 2004)

**Table 11** Bates pricing comparison

| Strike | Market Price | Bates Model Price | Absolute Error |
|---|---|---|---|
| **235** | 16.05 | 15.81 | 0.24 |

The results has excellent local fit and confirms calibration is accurate for practical pricing use *(Lewis, 2001).* We used the put call parity value and obtained the following heston parameters,

**Table 12** Recalibrated Heston parameters

| Parameter | Symbol | Calibrated Value |
|---|---|---|
| **Mean reversion** | $(\kappa_v)$ | 6.011 |
| **Long-run variance** | $(\theta_v)$ | 0.2027 |
| **Vol-of-vol** | $(\sigma_v)$ | 0.020265 |
| **Correlation** | $(\rho)$ | −0.04184 |
| **Initial variance** | $(v_0)$ | 0.0027255 |

**Table 13** Sanity-check pricing (alt dataset)

| Price Type | Value |
|---|---|
| **Market price** | 14.75 |
| **Heston-model price** | 14.61 |

**Table 14** Alt jump calibration

| Parameter | Meaning | Calibrated Value |
|---|---|---|
| $(\lambda)$ | Jump intensity | ~0.00 |
| $(\mu_J)$ | Mean relative jump size | −0.50% |
| $(\sigma_J)$ | Jump volatility | ~0 |

**Table 15** Interpretation (alt calibration)

| Result | Meaning |
|---|---|
| $(\lambda \approx 0)$ | Market does not support frequent jumps at 60-day maturity |
| Very small $(\sigma_J)$ | Little variation in jump sizes |
| Slightly negative $(\mu_J)$ | Low volatility (close to zero) |

The Heston model's price (14.61) closely matches the market price (14.75), showing that stochastic volatility alone explains most of the option's value. When jumps are added, the estimated parameters, jump intensity λ≈0, jump volatility σJ≈0, and mean jump size μJ≈−0.5%, suggest that the market does not expect frequent or large jumps over a 60-day horizon. The small negative μJ mild downside risk but no strong evidence of crash-like events. Overall, the calibration indicates a stable market regime where continuous diffusion dominates and jump risk is negligible (Bates, 1996; Pan, 2002; Cont & Tankov, 2004).

### 4.2.2 Bates Model Calibration via Carr Madan (2001)

We obtain calibrated Bates (1996) parameters in table 16.

**Table 16** Bates parameters (FFT)

| Parameter | Symbol | Calibrated Value |
|---|---|---|
| **Mean reversion** | $(\kappa_v)$ | 15.643 |
| **Long-run variance** | $(\theta_v)$ | 0.15858 |
| **Vol-of-vol** | $(\sigma_v)$ | 0.000004784 |
| **Correlation** | $(\rho)$ | −0.00231 |
| **Initial variance** | $(v_0)$ | 0.0485 |
| **lambda** | | 3.199e-15 |
| **mu** | $\mu$ | -0.573 |
| **delta** | $\delta$ | 0.0000809 |

These values suggest very fast mean-reversion, near-constant variance (σv essentially 0) and weak leverage effect (ρ close to 0). The jump intensity is also approximately zero. Looking at both the Lewis and Carr Madan approaches, we can observe that the calibrated parameters are not exactly the same but are very close to each other.

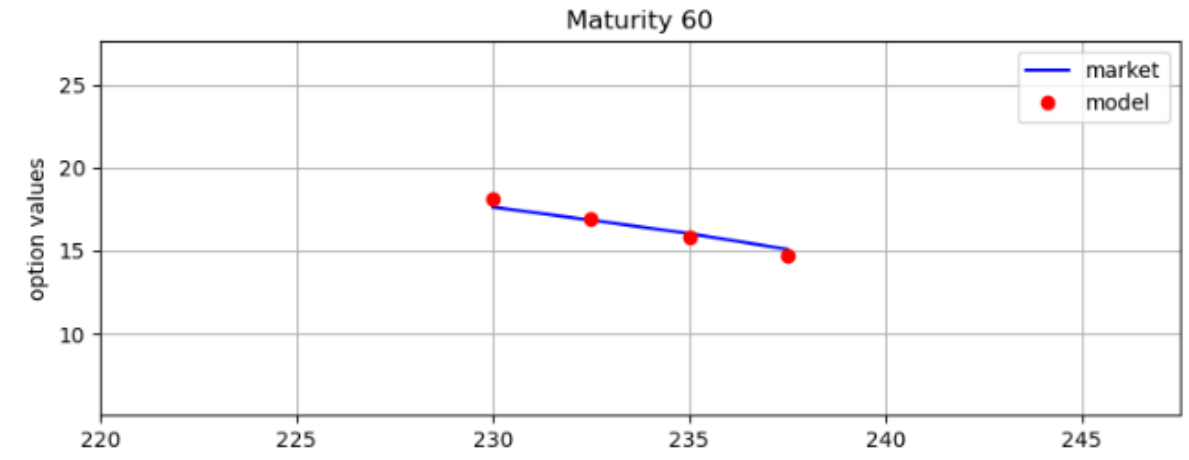

**Figure 14** Bates model fit (Carr-Madan)

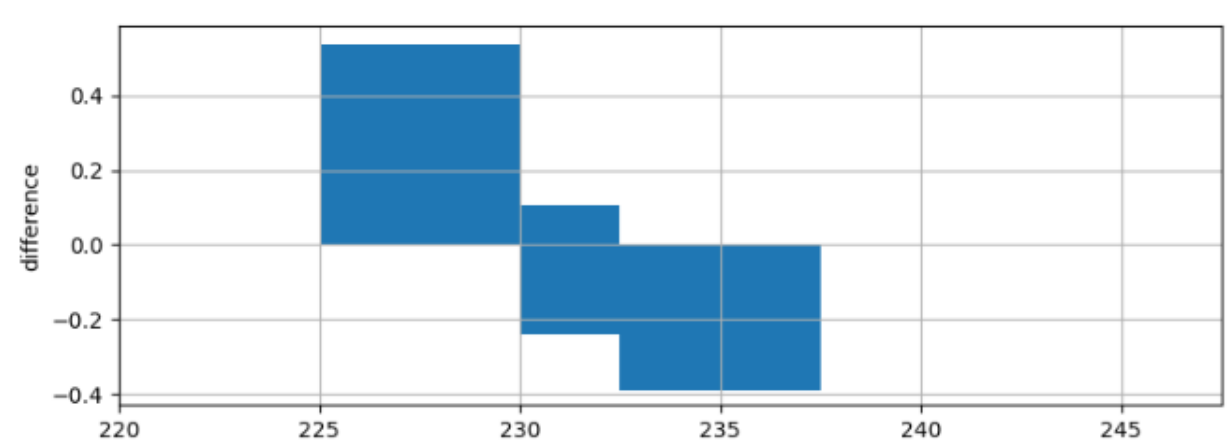

**Figure 15** Bates model fit (Carr-Madan FFT )

As expected, the model fit to market prices is similar to that of Lewis because of the near zero deviation of the parameters for both approaches

### 4.2.3 Pricing 70-Day 95% Moneyness Put

Using K=0.95$S_0$, T=70/250.
Put price: $P = Ke^{-rT}(1 - P_2) - S_0(1 - P_1)$
*(Heston 1993; Bates 1996).*

We obtained the following results,

1. n_paths = 120000, n_steps =70, elapsed =2.95s
2. Strike (K) = 221.2550
3. Fair (discounted) Asian put price = 1.987091
4. Std. error = 0.017402
5. 95% CI = [1.952983, 2.021200]
6. Client price (with 4% fee) = 2.066575
7. Fair (discounted) Asian Put price = 1.99
8. Std error = 0.02
9. 95% CI = [1.95, 2.02]
10. Client price (4% fee) = 2.07

### 4.3 Step 3 – CIR (1985) Interest-Rate Model

### 4.3.1 Calibration to Euribor Term Structure

We calibrate the CIR short-rate model to the Euribor curve (1w–12m), constructing weekly points by cubic-spline interpolation and solving a constrained least squares that enforces economic plausibility on the long-run level θ ≤ 5%, preserves positivity (Feller), and lets κ and σ flexibly fit the curve. (Cox, Ingersoll & Ross, 1985; Brigo & Mercurio, 2006). Objective (discount-factor fit):

$$\min_{\kappa,\theta,\sigma,r_0} \frac{1}{N}\sum_{i=1}^{N}\left(P_{CIR}(0,T_i) - P_{mkt}(0,T_i)\right)^2$$

with the CIR zero-coupon bond pricing formula:

$$P(0,T) = A(0,T)\, e^{-B(0,T) r_0}, \quad \gamma = \sqrt{\kappa^2 + 2\sigma^2}$$

$$B(0,T) = \frac{2\,(e^{\kappa T} - 1)}{(\kappa + \gamma)(e^{\kappa T} - 1) + 2\gamma},$$

$$A(0,T) = \left[\frac{2\gamma\, e^{(\kappa+\gamma)T/2}}{(\kappa + \gamma)(e^{\kappa T} - 1) + 2\gamma}\right]^{\frac{2\kappa\theta}{\sigma^2}}$$

(Cox, Ingersoll & Ross, 1985; Brigo & Mercurio, 2006)

The calibrated parameters (annualized) are:

**Table 17** Interpretation of Final CIR Calibration

| Parameter | Value | Meaning / Interpretation |
|---|---|---|
| ( $\kappa$ ) (speed of mean reversion) | 2.0000 | Rates revert quickly to their long-run level |
| ( $\theta$ ) (long-run equilibrium EURIBOR) | ≈ 4.22% | Represents market expectation of stable long-term rates around 4-4.5% |
| ( $\sigma$ ) (volatility) | 0.4110 | Moderate interest-rate uncertainty |
| ( $r_0$ ) (initial short rate) | 0.404% | Initial rate aligned with the short-end of the EURIBOR curve |
| Fit Error | RMSE = 3.65 bps | Excellent replication of the observed term structure |

Notes:

1. *The Feller condition $2\kappa\theta > \sigma^2$ is approximately satisfied under the calibrated parameters, ensuring rate positivity.*

Model Fit quality:

1. RMSE: 3.65 bp
2. MAE: 2.25 bp

Model checks:

Mean-reversion half-life is $t_{1/2}$ = ln2/$\kappa$ ≈ 0.35 years.

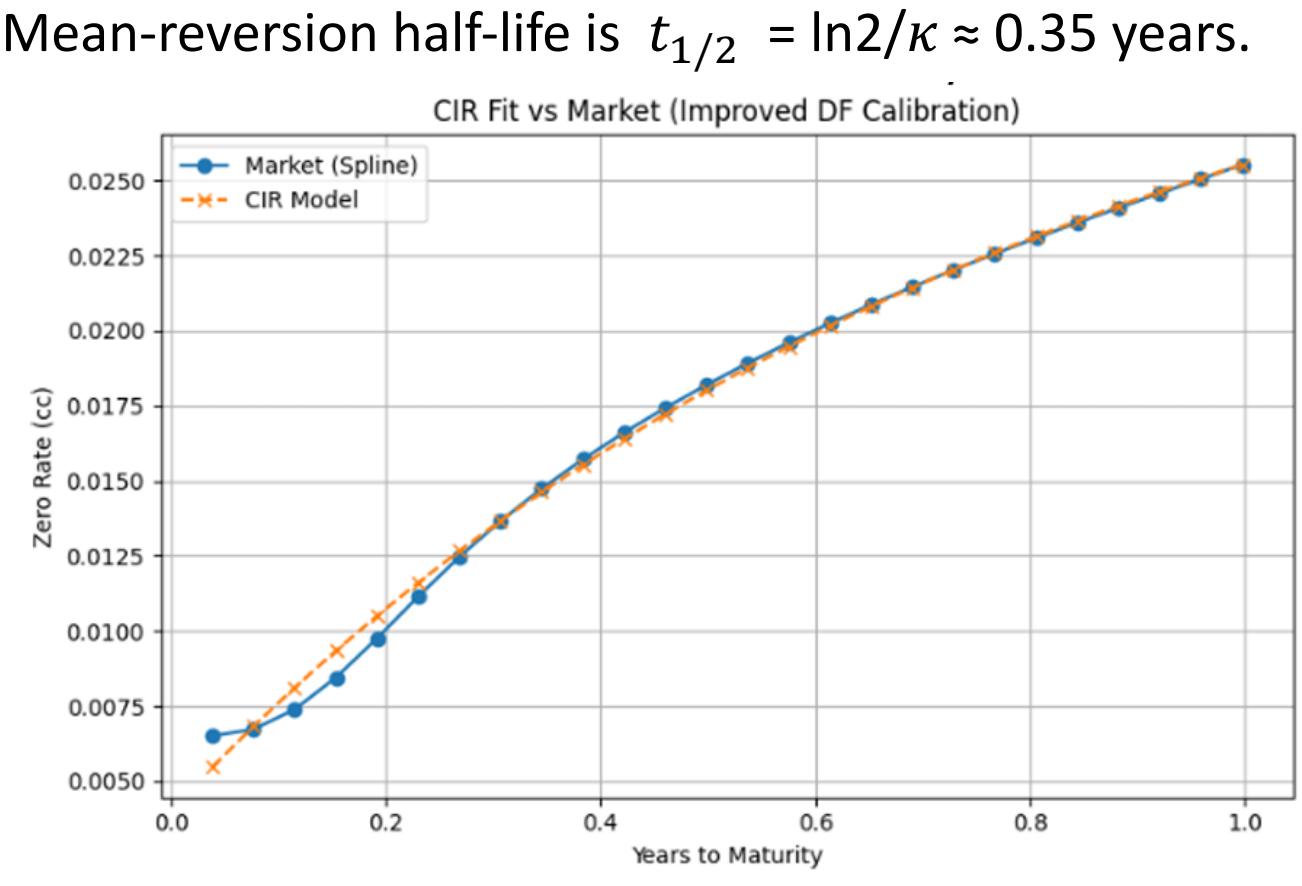

**Figure 16** CIR Fit vs Market Curve

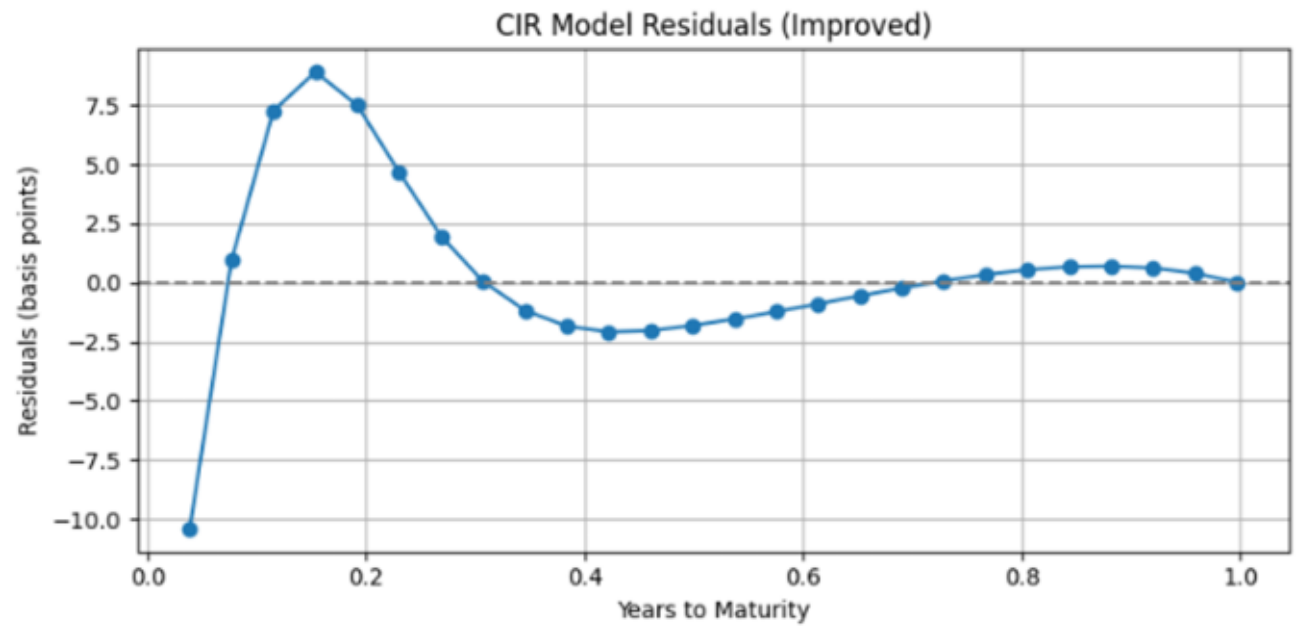

**Figure 17** CIR Model Residuals

Half-life is 0.35 years meaning deviations from θ decay by half in about 4 to 5 months, and front-end rates adjust fast toward equilibrium.

### 4.3.2 One-Year Monte-Carlo (100,000 paths) & Risk Insights

We simulated 100,000 daily paths under the calibrated CIR dynamics:

$$dr_t = \kappa(\theta - r_t)\, dt + \sigma\sqrt{r_t}\, dW_t$$

Risk-neutral simulation scheme (daily steps, $\Delta t$):

$$r_{t+\Delta t} = r_t + \kappa(\theta - r_t)\Delta t + \sigma\sqrt{r_t}\,\sqrt{\Delta t}\,\varepsilon_t, \quad \varepsilon_t \sim \mathcal{N}(0,1).$$

(Glasserman, 2004)

**Table 18** 12M Euribor distribution

| Statistic | Value |
|---|---|
| **Mean** | 3.73% |
| **Median** | 2.59% |
| **5th percentile** | 0.19% |
| **95th percentile** | 11.13% |

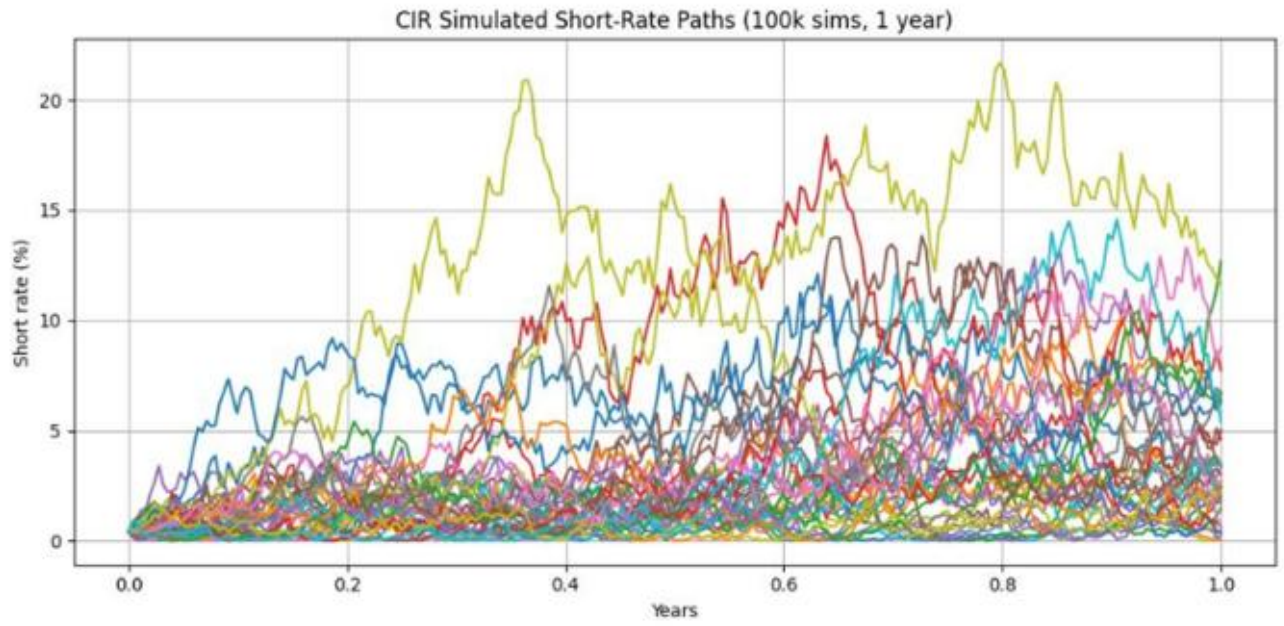

**Figure 18** CIR-simulated short-rate paths

This plot shows a wide variety of rate trajectories over the next year. Some simulations climb sharply, whereas others remain close to current levels. All remain above zero, which is expected since the CIR process enforces rate positivity and our parameters satisfy the Feller condition. The upward drift seen in many paths reflects the pull toward the long-run mean implied by our calibration (Cox et al., 1985; Brigo & Mercurio, 2006). This highlights meaningful uncertainty in the short-rate outlook, even over a 12-month horizon.

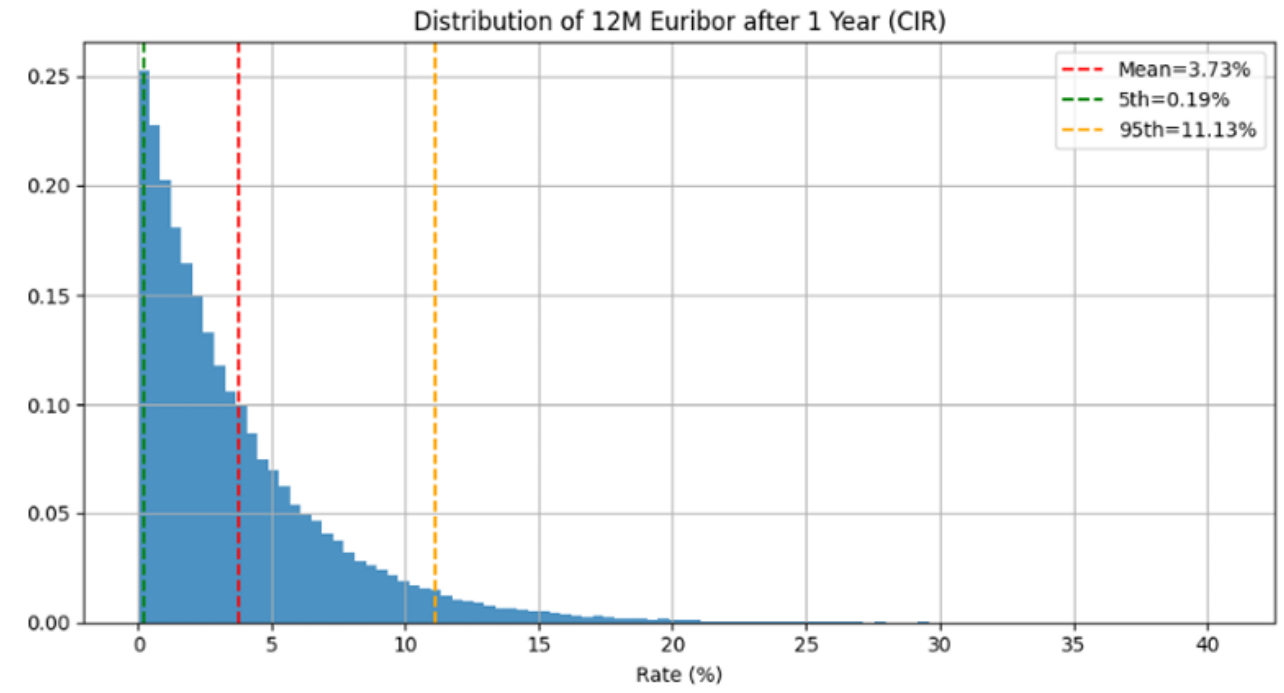

**Figure 19** Distribution of simulated 12M Euribor

The histogram is strongly right-skewed, which is typical of square-root rate models (Cox et al., 1985). While low-rate outcomes are possible, the model assigns noticeably more probability to higher rates, aligning with the current upward-sloping Euribor curve. The elevated mean relative to today's levels suggests the market anticipates further tightening but still carries wide uncertainty bands (Glasserman, 2004). To understand valuation consequences, we compute the discount factors implied by each path:

**Table 19 Discount factors vs flat curve**

| Statistic | Result |
|---|---|
| **Mean DF** | 0.9745 |
| **Flat-curve DF (today's 12M rate)** | 0.996 |
| **Difference** | −2.16% vs flat pricing |

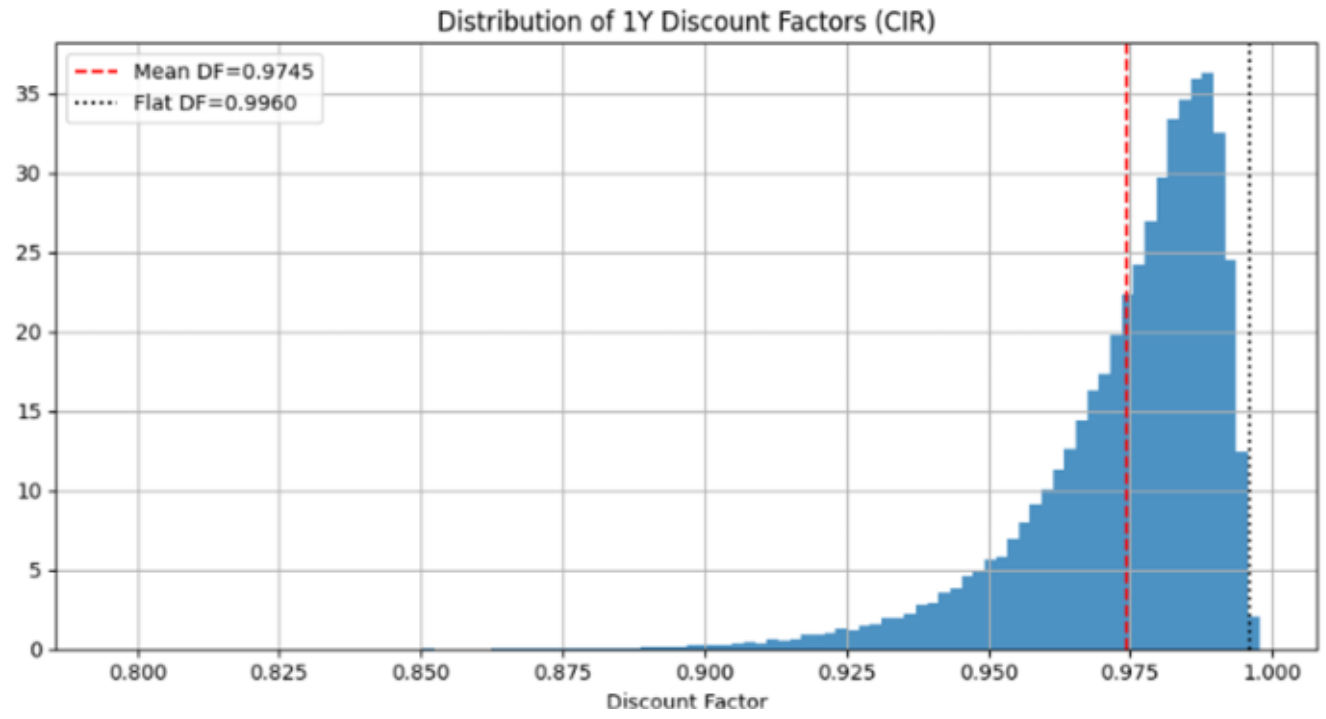

**Figure 20** Discount factor distribution

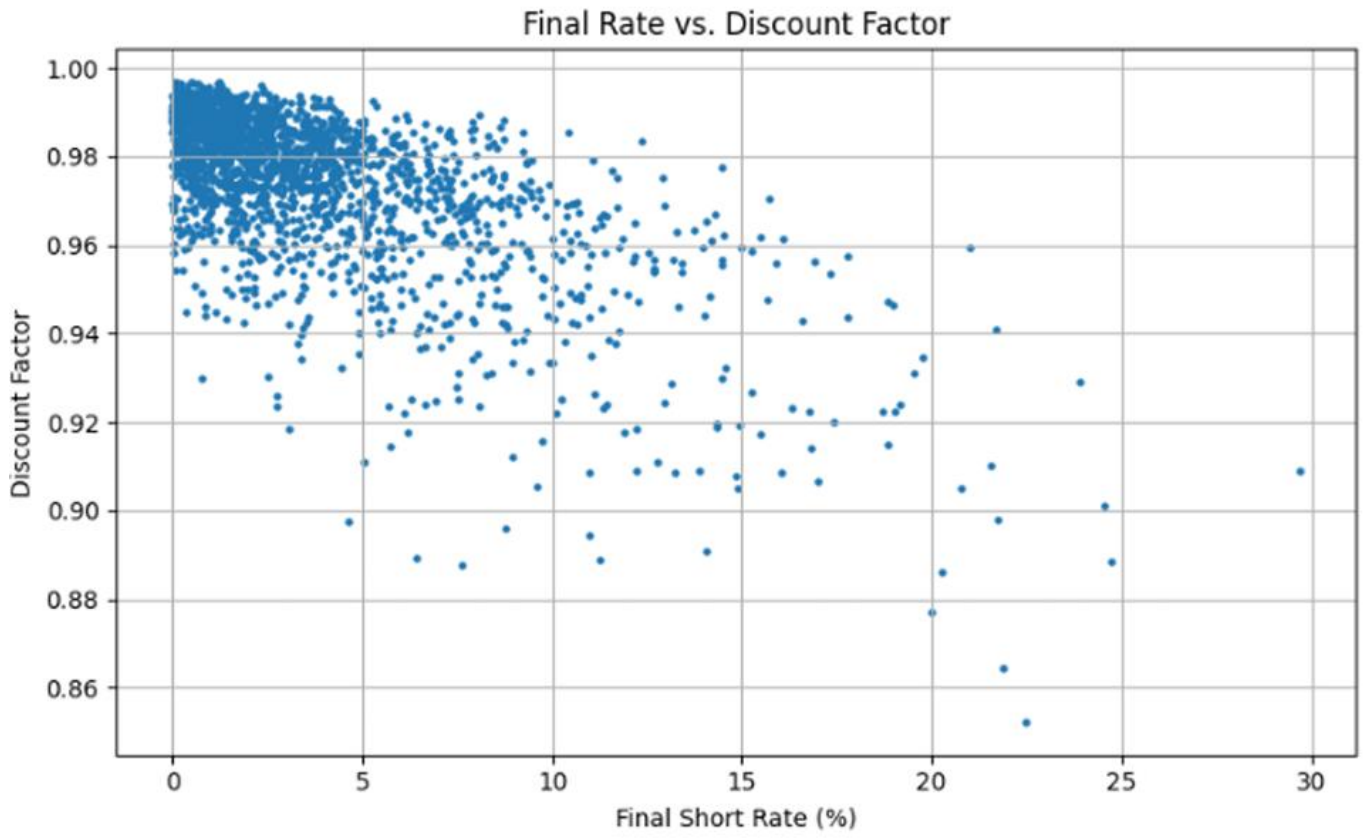

**Figure 21** Rate-discount factor scatterplot

The average discount factor is lower than the flat-rate benchmark. This means that using a flat curve would overvalue future cash flows by roughly 2.2%. The scatter plot illustrates the negative relationship expected under risk-neutral pricing: higher final rates correspond to stronger discounting. The spread of points across the chart shows how the exact discount depends not only on the final rate, but on the entire path taken to get there (Glasserman, 2004).

## 5. Results and Discussion

In this section we synthesize the insights from the calibration exercises and highlight the implications for model selection and pricing accuracy. The calibration results across the three modeling frameworks provide several insights that are valuable for pricing and risk analysis. The two Fourier-based methodologies used for the Heston model, Lewis (2001) and Carr-Madan (1999), returned almost identical parameter sets, which supports the stability of the estimation process. The volatility-of-volatility parameter collapsed to near zero across maturities, indicating that the market does not currently require strongly time-varying volatility to explain short-dated SM option prices. The correlation parameter appeared close to +1, which is unusual relative to the typical negative leverage effect observed in broader equity indices, where volatility tends to rise when prices fall (Heston, 1993; Gatheral, 2006). However, model–market price errors remained low, suggesting that the implied volatility surface for this particular stock is nearly flat over the examined maturities (Cont & Tankov, 2004).

Monte-Carlo pricing of the 20-day Asian call showed only a minor difference (less than one cent) between the Heston parameters calibrated by each Fourier approach, confirming that pricing outputs are robust to the choice of numerical transform (Glasserman, 2004; Hull, 2018). When extending the model to Bates (1996) for the 60-day calibration, the jump component did not activate materially: the jump intensity converged to a value effectively equal to zero and jump volatility was negligible. This outcome implies that market prices did not signal significant event-driven risk at this horizon, a result consistent with relatively calm market conditions, where stochastic volatility alone can capture the observed skew (Pan, 2002; Cont & Tankov, 2004).

The Bates pricing fit remained excellent, reinforcing the conclusion that discontinuous jumps are not required for valuation accuracy in this dataset. Pricing the 70-day put at 95% moneyness using the calibrated Bates parameters yielded a fair value close to USD 1.99, with a narrow confidence interval. This stability further confirms that adding a jump component does not materially change pricing at this maturity.

The improved CIR calibration produced a very strong match to the market term structure, with discount-factor RMSE of around 3.6 basis points. A moderately high mean-reversion speed ($\kappa \approx 2.0$) combined with a long-run equilibrium rate of approximately 4.2% leads to a half-life of about 0.35 years, indicating that shocks to short-term rates tend to subside within a few months (Cox et al., 1985; Brigo & Mercurio, 2006). The model also satisfies the Feller condition, ensuring strictly positive paths, which is essential for arbitrage-free interest-rate modeling (Cox et al., 1985). Overall, the fitted parameters reflect a market environment in which EUR money-market rates are expected to stabilize but remain higher than initial levels, consistent with current macroeconomic uncertainty.

## 7. Conclusion

This study develops an integrated stochastic-modeling framework combining the Heston, Bates, and CIR models to evaluate short- and medium-maturity derivative prices. Using SM Energy options with maturities of 15, 60, and 120 days, we show that Fourier-based calibration techniques, Lewis (2001) inversion and the Carr–Madan (1999) FFT, produce nearly identical Heston parameter estimates, confirming numerical robustness consistent with recent findings on calibration stability in stochastic volatility models (Agazzotti et al., 2025). For 60-day maturities, the Bates extension yields jump intensities close to zero, aligning with recent evidence that short-term implied-volatility structures can often be captured without substantial jump components (Agazzotti et al., 2025).

The CIR model calibrated to the Euribor term structure produces economically intuitive parameters and realistic forward-rate scenarios, consistent with recent analyses of stochastic-rate environments in option pricing (Jeon & Kim, 2025). These results collectively indicate that continuous stochastic volatility dominates short-maturity price dynamics, while stochastic interest rates become increasingly important beyond one year. Overall, the integrated modeling pipeline provides a flexible and empirically grounded approach for derivative pricing across maturities, situating our findings within the broader literature on calibration robustness and multi-factor risk modeling (Yoo, 2025).

## 7. Model Limitations

Despite its strong empirical performance, the modeling framework has several limitations. First, parameter stability is examined for a single underlying asset; broader cross-sectional validation may reveal different behaviors, particularly under stressed markets (Pan, 2002; Agazzotti et al., 2025). Second, the Bates model assumes Normally distributed log-jumps, which may be restrictive given empirical evidence of heavy-tailed jump distributions (Cont & Tankov, 2004). Third, the CIR model enforces strictly positive interest rates, which may limit applicability in low- or negative-rate environments, a limitation highlighted in recent interest-rate modeling studies (Jeon & Kim, 2025). Fourth, calibration uses mid-quotes without incorporating liquidity premia or bid-ask uncertainty. Finally, the analysis assumes a risk-neutral measure without assessing the role of alternative pricing kernels or incomplete-market frictions.

## 8. Future Research

Several extensions offer opportunities for future research:

1. Incorporating more flexible jump structures, such as double-exponential jumps, as examined in Agazzotti et al. (2025), or tempered-stable processes—may improve the fit for markets with pronounced skew or kurtosis.
2. Adopting joint calibration approaches that penalize both volatility-surface and variance-term-structure misalignment, following Yoo (2025), could further stabilize model parameters across maturities.
3. Extending the CIR component to multi-factor affine models may better capture shifts in the yield curve, especially in environments with structural interest-rate changes (Jeon & Kim, 2025).
4. Incorporating stochastic correlation or rough volatility dynamics could better capture high-frequency features of the spot–volatility relationship. Finally, extending the framework to multi-asset derivatives or portfolio-level risk applications would provide a broader test of model robustness in higher dimensions.